\documentstyle [12pt] {article}
\title{ \Large \bf Naked Singularities as Possible Candidates for
Gamma-ray Bursters}
\author
{Sandip K. Chakrabarti and Pankaj S. Joshi\\
Tata Institute of Fundamental Research, Homi Bhabha Road, Colaba,\\
Bombay, 400005, INDIA\\}
\begin{document}
\baselineskip 22pt
\maketitle

\begin{abstract}

Naked singularities appear  naturally
in dynamically evolving  solutions of Einstein equations
involving gravitational collapse
of radiation, dust and perfect fluids, provided the rate of
accretion is less than a critical value.
We propose that the gamma-ray bursters (GRBs) are examples of these
naked singularity solutions. For illustration, we show that
according to solutions involving spherically symmetric
collapse of pure radiation field, the energy $E_\gamma$ and
the observed duration $\Delta t_o$ of a GRB
should satisfy, $\frac{E_\gamma}{\Delta t_o}
\leq 4.5 \times 10^{58} \ f_\gamma $ erg sec$^{-1}$, $f_\gamma$
being the fraction ($10^{-2}$  to $10^{-3}$)
of energy released as gamma rays. All the presently observed
GRBs satisfy this condition; those satisfying the condition
close to equality must necessarily be of cosmological origin with the
red-shift factor $z$ not exceeding $\sim 2-10$ depending on exact
observed flux.

\end{abstract}

\newpage

Gamma-ray bursts (GRBs), both weak and strong, are believed to be
isotropically distributed as revealed by PVO and BATSE experiments$^{1,2}$.
These observations strongly argue for their cosmological origin$^{3-5}$.
The bursts have typical flux of $10^{-5}$ to $2 \times
10^{-4}$erg cm$^{-2}$ with the rise time as low as $10^{-4}$s and the
duration of burst from $10^{-2}$s to $10^3$s.
The origin of GRBs is not yet known. Speculations involve merger of binary
neutron stars$^{5-6}$ and capture of neutron stars by black holes$^7$.

In the present {\it letter}, we propose that the GRBs could be the
naked singularity solutions of Einstein's equations describing collapse
of radiation, dust or matter shells.
Recently, a large number of such solutions have been proposed$^{8-18}$.
The nature of such a naked singularity has been analyzed in detail and
it has been shown $^{8-13}$ that the space-time curvatures and gravitational
tidal forces grow very strongly in the vicinity of these singularities
which turn out to be strong curvature singularities in a very powerful
sense. Hence, it appears that for several reasonable equations of state
satisfying the positivity of energy, a strong curvature
naked singularity may be formed in the
space-time as a result of the gravitational collapse.
During the collapse, such naked singularities could emit powerful bursts
of radiation visible to an external observer situated far away from the
sight of collapse.

To focus on a particular solution of such kind,
we consider the case of gravitational collapse of a
spherical shell of radiation at the center of the symmetry. The total mass
of the singularity grows from zero to a finite total of $M$ when the
final collapsing shell has arrived at the singularity$^{14}$.
A naked singularity forms at the origin $t=0,r=0$,
which is a two-sphere in a space-time, or a null surface when represented
in a Penrose diagram. The gravitational potentials
(in the units $c=1, \ G=1$) within the radiation
zone are described by a Vaidya metric and are given by,
$$
g_{uu}= -\left(1-\frac{2m(u)}{r}\right),\   \  g_{ur}=1,
\ g_{\theta\theta}=r^2, \ g_{\phi\phi}=r^2 \ sin^2 \theta
\eqno{(1)}
$$
where $u$ is the advanced time given by $u=t+r$ and all other metric
components vanish.
After the completion of the collapse, the solution settles
to an external Schwarzschild geometry with a singularity of mass $M$ at the
center. It has been shown$^{8,9,16}$ that the naked
singularity at the center forms independently of the details of exact
functional form of the mass function. For the purpose of illustration,
it is convenient to choose a linear form for the mass function, say,
$m(u)=\lambda u$. Here, $\lambda$ is a constant defining the accretion rate
of the collapse at $r=0$, which in CGS unit is given by,
$$
\frac  {d m}{d t} |_{r=0} = \frac{ \lambda c^3}{G} \sim \Lambda \times
10^{38} {\rm gm\ sec}^{-1}.
\eqno{(2)}
$$
with $\Lambda=4 \lambda$.
It was shown$^{11-12}$ that the occurrence of the naked singularity depends
on the rate of accretion $\Lambda$. For $\Lambda \leq \frac{1}{2}$, the
collapse
produces naked singularity, but for $\Lambda >\frac{1}{2}$, black hole
solutions are formed.
We now assume that gamma ray luminosities from GRBs should not exceed
the rate of collapse of the radiant energy.
{}From the linear dependence of $m(u)$ on $u$, it then follows that
for a naked singularity solution,
$$
\frac{E_\gamma}{\Delta t_o} \leq
4.5 \times 10^{58} f_\gamma {\rm erg\ sec}^{-1}.
\eqno{(3)}
$$
Here, $f_\gamma$ denotes the conversion
fraction of the collapsing energy  which
is received as the gamma radiation and $\Delta t_o$ is the duration
of observation of the burst. From eqn. (3) one readily observes that
a collapse could correspond to conversion of a $20M_\odot$ object in
about a millisecond as measured on earth!
On the other hand, from the observational data$^1$, we require that,
$$
\frac{{E_\gamma}|_{obs}}{ 4\pi d_L^2}= 10^{-5} {\rm \ to\ }
2 \times 10^{-4} {\rm erg\ cm}^{-2}
$$
where, $d_L$ is the luminosity distance of the  source, $d_L=2R_0 (1+z  -\sqrt
{1+z})$, with $R_0=c/H_0=
3 \times 10^3$ h$_{\rm 100}^{-1}$ Mpc. Here  $h_{\rm 100}$
is the Hubble's constant ($H_0$)in units of $100$ km sec$^{-1}$ Mpc$^{-1}$,
and $z$ denotes the redshift of the object due to the expansion of the
universe.
Thus, ${E_\gamma}|_{obs}=4.1 \times 10^{52}$ to $8.1 \times 10^{53}
(1+z-\sqrt{1+z})^2$ h$_{100}^{-2}$erg. If we assume $f_\gamma=10^{-3}$
and $\Delta t_o=10^{-2}$s, it is easy to verify that the
condition (3) is readily satisfied for all the observed GRBs
unless they are very distant ($z \geq 1.25-4.6$). Instead, if we had chosen
$f_\gamma=10^{-2}$, eqn. (3) would have been satisfied unless $z \geq
3.5-13.5$. In this context, we would like to recall that in the models of GRBs
involving merger of neutron stars it is customary to choose$^1$
$f_\gamma = 10^{-2}$ to $10^{-3}$.

In this {\it letter} we have demonstrated that
the collapse of spheres filled with radiation fluid could produce bursts of
energetic radiation. The energy is expected to be in the band
$\sim \alpha m_p c^2 erg\sim 100Mev$, where  $m_p$ is the mass of the proton,
and $\alpha \sim=0.1$ depends on the
red-shift factor and the efficiency of conversion of accretion
energy into radiation. The naked singularities
could thus be possible basis for the gamma-ray bursters. The existence
of a cut-off in the accretion rate $\lambda$ enabled us to
separate the naked singularity solutions from those which produce
black holes at cosmological distances. We show that depending upon the
exact observed flux and the efficiency, GRBs should not be
located beyond $z \sim 2-10$. This could be a signature of the proposed
mechanism. In more realistic
collapse of dust$^{13,15}$, aspherical collisionless gas$^{18}$,
and perfect fluids$^{10,17}$, one also has similar parameters
which produce naked singularity solutions in a certain range of their
values. An interesting property of some of these solutions is that
through a single collapse, separate singularities might be developed
at different times. These solutions could be relevant to explain the
repeated bursts$^1$ which are observed. Some repeaters could also be
due to quasi-periodic oscillations induced during the collapse process.
In passing, we may remark that the time-variabilities of
emitted radiation from such
astrophysical systems should not be limited by the light-crossing
time of the Schwarzschild radius, as is currently assumed in describing
Active Galaxies and Quasars.
Detailed behavior of these solutions in the context of
observed astrophysical processes, such as line emissions from GRBs, etc.
will be dealt with elsewhere.

We wish to thank Prof. S.M. Chitre for reading the manuscript
and making many valuable suggestions.

\vspace {2.0cm}

\centerline {\Large References}

\noindent 1. Higdon, J.C. \& Lingenfelter, R.E., {\it Ann. Rev. Astron.
Astrphys
   },
{\bf 28}, 401-436, (1990)\\
2. Meegan, et al., {\it Nature}, {\bf 355}, 143-146, (1992).\\
3. Paczy\'nski, B., {\it Acta Astron.}, {\bf 41}, 257-263, (1991)\\
4. Mao, S. \& Paczy\'nski, B., {\it Astr. J.}, {\bf  388}, L45-L48, (1992)\\
5. Piran, T., {\it Astr. J.}, {\bf 389}, L45-L49, (1992)\\
6. Phinney, E.S., {\it Astr, J.}, {\bf 380}, L17-L20, (1991)\\
7. Carter, B., {\it Astr. J.}, {\bf 391}, L67-L70,  (1992)\\
8. Joshi P.S. and Dwivedi I.H.,{\it Gen. Rel. Grav.},{\bf 24}, 129-137,
(1992)\\
9. Joshi P.S. and Dwivedi I.H. , {\it Phys. Rev. D}, {\bf 45}, 2147-2150,
(1992)\\
10. Joshi P.S. and Dwivedi I.H., {\it Commun. Math. Phys.}, {\bf 146},
333-343, (1992)\\
11. Dwivedi I.H. and Joshi P.S.,{\it Class. Quantum Grav.},{\bf 6},1599-1607,
(1989)\\
12. Dwivedi I.H. and Joshi P.S., {\it Class. Quantum Grav.}, {\bf 8},1339-1349,
(1991)\\
13. Dwivedi I.H. and Joshi P.S., {\it Class. Quantum Grav.}, {\bf 9},L69-L75,
(1992)\\
14. Papapetrou in `A Random Walk in Relativity and Cosmology'
(eds. N. Dadhich et al), Wiley Eastern, New Delhi, (1985)\\
15. Eardley D.M. and Smarr L., {\it Phys. Rev.}, {\bf D19}, 2239,
(1979)\\
16. Lake K., {\it Phys. Rev.}, {\bf D43}, 1416, (1991)\\
17. Ori A. and Piran T., {\it Phys. Rev.}, {\bf D42}, 1068, (1990)\\
18. Shapiro S.A. and Teukolsky S.A., {\it Phys. Rev. Lett}, {\bf 66},994,
(1991)\\
\end{document}